\documentclass[reprint, pra, onecolumn,floatfix]{revtex4}
\usepackage[latin1]{inputenc}
\usepackage{enumerate}
\usepackage{amsfonts}
\usepackage{amssymb}
\usepackage{amsmath,amsbsy}
\usepackage{dcolumn}
\usepackage{bm}
\usepackage{graphicx, graphics}
\usepackage{color}

\newcommand{\pac}[1]{ \left\{ #1 \right\} }
\newcommand{\pap}[1]{\left( #1 \right)}

\newcommand{\Jo}{\hat{J}}

\newcommand{\rhoo}{\hat{\rho}}
\newcommand{\bra}[1]{\left\langle #1 \right\vert}

\newcommand{\ket}[1]{\left\vert #1 \right\rangle}
\newcommand{\tr}[1]{\mathrm{tr}\left\{ #1 \right\}}
\newcommand{\ee}{\mathrm{e}}

\newcommand{\lam}{\lambda}
\newcommand{\ome}{\omega}
\newcommand{\eps}{\epsilon}

\begin{document}
\title{Quantum Hysteresis in Coupled Light-Matter Systems}
\author{F. J. G\'omez-Ruiz$^{1}$}
\email{ fj.gomez34@uniandes.edu.co}
\author{O. L. Acevedo$^{2}$}
\author{L. Quiroga$^{1}$}
\author{F. J. Rodr\'iguez$^{1}$}
\author{N. F. Johnson$^{3}$}
\affiliation{$^1$Departamento de Física, Universidad de los Andes, A.A. 4976, Bogotá D. C., Colombia}
\affiliation{$^2$JILA, University of Colorado, Boulder, CO 80309, U.S.A.}
\affiliation{$^3$ Department of Physics,  University of Miami, Coral Gables, FL 33124, U.S.A.}
\date{\today}
\begin{abstract}
We investigate the non-equilibrium quantum dynamics of a canonical light-matter system, namely the Dicke model, when the light-matter interaction is ramped up and down through a cycle across the quantum phase transition. Our calculations reveal a rich set of dynamical behaviors determined by the cycle times, ranging from the slow, near adiabatic regime  through to the fast, sudden quench regime. As the cycle time decreases, we uncover a crossover from an oscillatory exchange of quantum information between light and matter that approaches a reversible adiabatic process, to a dispersive regime that generates large values of light-matter entanglement. The phenomena uncovered in this work have implications in quantum control, quantum interferometry, as well as in quantum information theory.
\end{abstract}
\maketitle
\section{Introduction}
In the last two decades, there have been several breakthroughs in the experimental realization of systems that mimic specific many-body quantum models~\cite{Han2013}. This is especially true in systems involving aggregates of real or artificial atoms in cavities and superconducting qubits~\cite{Ciuti2010,Marquadt2011}, as well as trapped ultra-cold atomic systems~\cite{bloch2012nat,schneider2012rpp,georgescu2014rmp}. These advances have stimulated a flurry of theoretical research on a wide variety of phenomena exhibited by these systems, such as quantum phase transitions (QPTs)~\cite{sachdev, aeppli}, the collective generation and propagation of entanglement~\cite{amico2002nature, wu2004prl, RomeraPLA2013, Reslen2005epl, Juan2010pra, Acevedo2015NJP, AcevedoPRA2015}, the development of spatial and temporal quantum correlations~\cite{sun2014pra,FernandoPRB2016}, critical universality~\cite{Acevedo2014PRL}, and finite-size scalability~\cite{Vidal2006, Oct_PS2013, CastanoPRA2011, CastanoPRA2012}. All of these topics have implications for quantum control protocols which are in turn of interest in quantum metrology, quantum simulations, quantum computation, and quantum information processing~\cite{Gernot2006,Rey2007,Dziarmaga2014AP,Hardal_CRP2015,NiedenzuPRE2015}. \\
\\
Future applications in the area of quantum technology will involve exploiting -- and hence fully understanding -- the {\em non-equilibrium} quantum properties of such many-body systems. Radiation-matter systems are of particular importance: not only because optoelectronics has always been the main platform for technological innovations, but also in terms of basic science because the interaction between light and matter is a fundamental phenomenon in nature. On a concrete level, light-matter interactions are especially important for most quantum control processes, with the simplest manifestation being the non-trivial interaction between a single atom and a single photon~\cite{AgarwalPRL1984}. One of the key goals of experimental research is to improve both the intensity and tunability of the atom-light interaction~\cite{WillPRL2016, BegleyPRL2016}. Unfortunately the relatively weak dipolar coupling between an atom and an electromagnetic field makes it difficult to obtain a large light-matter interaction, even when atoms are constrained to interact with a single radiation mode in a cavity. In recent years, some condensed matter systems have offered an alternative to the traditional atom-cavity implementation. Clear spectroscopic evidence has recently been presented that a charged Josephson qubit coupled to a superconductor transmission line, behaves like an atom in a cavity~\cite{XiuPRA2016} and that the dipolar coupling between these systems is $3$ to $4$ orders of magnitude greater than that in atomic systems. This type of system, known as a superconducting qubit, enables the study of effective two-level atoms interacting with a quantized single-mode electromagnetic field~\cite{StefanoPRA2016, LijunPRA2016} and allows the exploration of new regimes of strong coupling~\cite{HerreraPRL2016}. Another very successful approach to obtain strongly interacting light-matter systems has emerged in experiments involving ultra-cold trapped atoms or ions~\cite{Baumann2010,islam2011nat}. In this case, discrete translational degrees of freedom (vibrational modes) emerging from the optical trap are used to couple the atoms to the radiation mode. Thanks to the extremely low temperatures, the light-matter coupling effectively becomes the dominant interaction, once again allowing the exploration of a wide variety of strong-coupling phenomena.\\
\\
One important consequence of such strong coupling, is that the atom-light interaction can effectively become all-to-all, in the sense that all atoms are equally coupled to the radiation. In this case, one of the simplest and yet richest scenarios involves instances in which the Dicke model (DM) is realized~\cite{DickePR1954}. One of the most striking and important features of the Dicke model is the fact that it exhibits a superradiant second order QPT in the thermodynamic limit~\cite{Lieb1973}. Despite more than sixty years of existence, this model has recently attracted renewed interest thanks to major experimental breakthroughs in terms of its realization and exploration~\cite{Baumann2010,KlinderPNC2015,GuerinPRL2016}. This has in turn fueled a surge in theoretical investigations of the DM, including further detailed proposals for its realization~\cite{WangPRA2016}. Regardless of this surge in theoretical interest, however, much of the focus has been on the DM's static properties or equilibration schemes, leaving many aspects of its non-equilibrium evolution as an open problem.\\
\\
In previous work, we had attempted to advance understanding of the DM's dynamics by exploring the effects of crossing the QPT using a tuned interaction, hence taking the system in a single sweep from a non-interacting regime into one where strong correlations within and between the matter and light subsystems play an essential role~\cite{AcevedoPRA2015,Acevedo2015NJP,Acevedo2014PRL}. Our previous analyses also revealed universal dynamical scaling behavior for a class of models concerning their near-adiabatic behavior in the region of a QPT, in particular the Transverse-Field Ising model, the DM and the Lipkin-Meshkov-Glick model. These findings, which lie beyond traditional critical exponent analysis like the Kibble-Zurek mechanism~\cite{ZurekNAT1985, KibbleJOP1976} and adiabatic perturbation approximations, are valid even in situations where the excitations have not yet stabilized -- hence they provide a time-resolved understanding of QPTs encompassing a wide range of near adiabatic regimes. \\
\\
In this work, instead of a single crossing of the QPT, we analyze the effects of driving the system through a round trip across the QPT, by successively ramping up and down the light-matter interaction so that the system passes from the non-interacting regime into the strongly interacting region and back again. We restrict ourselves to the case of a closed DM such that a description of the temporal evolution using unitary dynamics is sufficient. Depending on the time interval within which the cycle is realized, we find that the system can show surprisingly strong signatures of \emph{quantum hysteresis}, i.e. different paths in the system's quantum state evolution during the forwards and backwards process, and that these memory effects vary in a highly non-monotonic way as the round-trip time changes. The adiabatic theorem ensures that if the cycle is sufficiently slow, the process will be entirely reversible. In the other extreme, where the round-trip ramping is performed within a very short time, the total change undergone by the system is negligible. However in between these two regimes, we find a remarkably rich set of behaviors.\\
\\
This paper is organized as follows: Section~\ref{DickeModel} describes the Dicke Model (DM) and discusses its quantum phase transition (QPT). Section~\ref{Hyste} describes the quantum hysteresis process of the DM and introduces the Landau-Zener model as a simple two-level approximation for understanding cyclical crossing of quantum critical points. Section~\ref{results} presents and discusses the results of quantum hysteresis that we uncover in the DM. Our focus is on two main quantities: the ground state fidelity and the von Neumann entropy. These quantities provide a perspective that is of interest to adiabatic quantum control in the former case, and quantum information theory and thermodynamics in the latter case. Section~\ref{conclusions} presents our final remarks.

\section{Dicke Model (DM) and its Quantum Phase Transition (QPT)}\label{DickeModel}

The DM of a many-body light-matter interacting quantum system, features a set of $N$ identical two-level systems (commonly referred to as qubits) each of which is coupled to a single radiation mode. It can be described by the following microscopic Hamiltonian:
\begin{equation}\label{hdic}
\hat{H}(t)=\frac{\epsilon}{2}\sum_{i=1}^{N}\hat{\sigma}_{z}^{i} + \omega \hat{a}^{\dagger}\hat{a} +\frac{\lambda(t)}{\sqrt{N}}\left(\hat{a}^{\dagger}+\hat{a}\right)\sum_{i=1}^{N}\hat{\sigma}_{x}^{i}\:.
\end{equation}
We purposely avoid common approximations such as the rotating-wave approximation. Here $\sigma_{\alpha}^{i}$ denotes the Pauli operators for qubit $i$ $\left(\alpha=x,z\right)$;  $\epsilon$ and $\omega$ represent the qubit and field transition frequencies respectively; and $\lambda(t)$ represents the strength of the radiation-matter interaction at time $t$ which can be varied over time. In many situations such as those we consider here in our work, the dynamics of the DM do not require the consideration of the entire $2^{\otimes N}\otimes \mathbb{N}$ dimensional Hamiltonian. Instead, $SU(2)$ collective operators $\hat{J}_{\alpha}=\frac{1}{2}\sum_{i=1}^{N}\hat{\sigma}^{i}_{\alpha}$ can be used. The Hamiltonian~\eqref{hdic} can then be written in the following form:
\begin{equation}\label{Hdicke}
\hat{H}(t)=\epsilon \hat{J}_{z} + \omega \hat{a}^{\dagger}\hat{a} +\frac{2\lambda(t)}{\sqrt{N}}\hat{J}_{x}\left(\hat{a}^{\dagger}+\hat{a}\right)\ \ .
\end{equation}
Here, operators $\hat{J}_{\alpha}$ act on the totally symmetric manifold, also known as the {\em Dicke manifold}, where the operator $\hat{{\bf J}}^{2}=\sum_{\alpha}\hat{J}_{\alpha}^{2}$ is a good quantum number (a constant of motion) with eigenvalue $J\left(J+1\right)$, and $J=N/2$. The parity operator $\hat{P}=\ee^{\Jo_z+\hat{a}^{\dagger} \hat{a}-N/2}$ is another conserved quantity of the model which also has the advantage of commuting with $\hat{{\bf J}}^{2}$. For all the numerical results in this paper, we consider the case of resonance between the qubits and the radiation frequency, i.e. $\epsilon=\omega=1$ in Eq.~\eqref{hdic}.
\begin{figure}[h!]
\begin{center}
\includegraphics[scale=0.93]{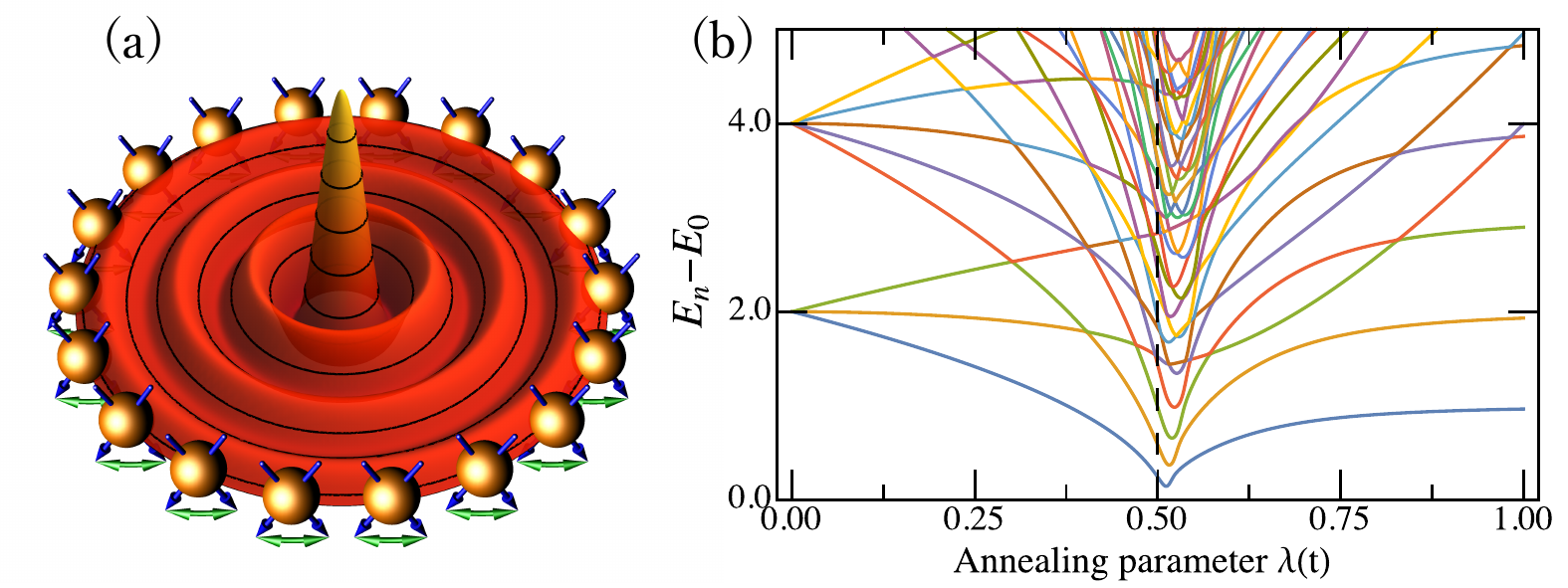}
\caption{\label{Espectro} {\bf (a)} Illustration of the Dicke model (DM) network of spins, or equivalently qubits. The electromagnetic field mode mediates the all-to-all interaction among these qubits. {\bf (b)} Energy spectrum of the DM as a function of the interaction parameter $\lam(t)$ for a finite number of qubits $N=257$. The numerical results in this paper are generated under the condition of resonance between the qubits and the radiation frequency: $\epsilon=\omega=1$ in Eq.~\eqref{hdic}. All energies are measured with respect to the ground state energy $E_{0}$. In the thermodynamic limit (TL), a second order QPT occurs at value $\lam_c= 0.5$ at which point  the energy gap vanishes. As can be seen in (b) for finite $N$, a finite-size version of the quantum critical point (QCP) with minimum gap arises near to $\lam_c$. Only the even parity sector of the model is depicted here.}
\end{center}
\end{figure}
The DM is deemed to achieve its thermodynamic limit (TL) when its matter subsystem size goes to infinity, i.e. when $N\to\infty$. Much of the DM's attractiveness as a theoretical model, lies in the emergence of a second-order QPT in this limit~\cite{Klauss, WangPRA, HioePRA}.  The QPT means that there is a significant change in the behavior of the DM's ground state when its parameters vary through a specific set of critical values. The phase-boundary is described by the equation $\lambda_{c} =\frac{\sqrt{\epsilon\omega}}{2}$. The scaling of the interaction parameter in Eq.~\eqref{hdic}  is introduced in order to have its phase transition correctly defined in terms of $\epsilon$ and $\omega$. When the coupling parameter $\lambda$ is above the critical value, the ground state of the system is characterized by a non-zero expectation value of its excitation operators or order parameters:
\begin{equation}
\bigl\langle \hat{N}_{b} \bigr\rangle\equiv \bigl\langle \hat{a}^{\dagger}\hat{a}\bigr\rangle \qquad \text{and} \qquad \bigl\langle \hat{N}_{q} \bigr\rangle\equiv \biggl \langle \hat{J}_{z} +\frac{N}{2}\biggr\rangle \:.
\end{equation}
When $\lambda < \lambda_{c}$, the order parameters are zero. Because of this, the region when $\lambda > \lambda_{c}$ is called the ordered or superradiant phase, while the region when $\lambda < \lambda_{c}$ is called the normal phase. Near the phase-boundary in the vicinity of this superradiant phase, there is a dependence of the order parameter as follows: $\bigl\langle \hat{N}_{b} \bigr\rangle\propto\left(\lambda -\lambda_{c}\right)$ and  $\bigl\langle \hat{N}_{q} \bigr\rangle\propto\left(\lambda -\lambda_{c}\right)^{1/2}$~\cite{EmaryPRE}. This power-law behavior is typical of second-order phase transitions where the critical exponents are characteristic of the universality class to which the model belongs. In the ordered quantum phase $\left(\lambda >\lambda_c\right)$, the $\mathbb{Z}_{2}$ symmetry related to parity is spontaneously broken, which originates from the fact that the TL ground state is two-fold degenerate corresponding to the two different eigenvalues of $\hat{P}$. Also at the QPT, the DM presents an infinitely-degenerate vanishing energy gap~\cite{EmaryPRE}, as shown in Fig.~\ref{Espectro}. \\
\\
This zero energy gap in the TL imposes an impossible barrier for dynamically crossing the phase boundary within any finite time while simultaneously maintaining the system in its ground state. For finite values of $N$, where the transition is replaced by an effective quantum critical point (QCP), there is a minimum finite gap that gets smaller and smaller as the system size $N$ grows. It also gets slightly shifted to higher values of $\lambda$ as compared to the TL case. Such scaling properties, i.e. dependence on size, have been extensively studied in the DM in the equilibrium situation~\cite{LambertPRL, VidalEPL, TengNJP, JarretEPL, ChenPRA, WangPRA2}.

\section{Quantum Hysteresis in the DM}\label{Hyste}
Our central objective in this paper is to address the effects of cyclically varying the radiation-matter interaction $\lambda(t)$ as a function of time. We will focus on a particularly simple, piecewise linear form for the time dependence:
\begin{equation}\label{ciclo}
\lambda(t)=\begin{cases}
\lambda_{1}+\frac{2\left(\lambda_2-\lambda_1\right)}{\tau}t,& t\leq \tau/2 \\
\lambda_{2}+\frac{2\left(\lambda_1-\lambda_2\right)}{\tau}t, & t>\tau/2 \:,
\end{cases}
\end{equation}
where $\lambda_1$ and $\lambda_2$ are respectively zero and one, and with $\ome=\eps=1$ in Eq.~\eqref{hdic}. Hence $\lambda(t)$ has a triangular profile (see inset of Fig.~\ref{Fide}{\bf (a)}). The slope of the two portions of the cycle $\pm\nu$ is characterized by a finite time $\tau$ such that $\nu= 2/ \tau$, where $\nu$ is known as the  {\it annealing velocity}. The strongly interacting regime is reached when $\lambda(t)\approx 1$. The particular choice of $\lambda(t)$ given by Eq.~\ref{ciclo}, implies that the QCP is crossed twice during the cycle, first when $t\approx\tau/4$, and second when $t\approx3\tau/4$. From Eq.~\eqref{Hdicke} and Eq.~\eqref{ciclo}, we get the full DM instantaneous state $\ket{\psi(t)}$ by numerically solving the time-dependent Schr\"{o}dinger equation. The initial state when $t=0$ and $\lam=0$ is the non-interacting ground state, namely $\ket{\psi(0)}=\ket{\psi_{GS}\pap{0}}=\bigotimes_{i=1}^{N}\ket{\downarrow}\otimes\ket{n=0}$ where both matter and light subsystems have zero excitations. All qubits are polarized in the state $\sigma_z=-1$, and the field is in the Fock state of zero photons. The DM solution is significantly harder to obtain. Nevertheless, we have used the fact that the total angular momentum  is a conserved quantity, therefore the dynamics will lie in the $J=N/2$ subspace. The success of this solution was tested by extending the truncation limit and then checking that the results do not change (convergence test). The most naive application of this solution to the DM would be to work with vectors of the form $\ket{m}_{z}\otimes \ket{n}$, where the first one is an eigenvector of $J_z$ in the subspace of even parity and the last one is a bosonic Fock state.\\
\\
To provide a first step toward understanding the complexity of the quantum hysteresis results that we generate for the DM, there exists a model that constitutes arguably the simplest version of what happens to a quantum system when it crosses a QCP driven by a time-dependent Hamiltonian. This is the Landau-Zener (LZ) model and we will spend the rest of this section reviewing its key properties concerning the probability that the system transitions out of its ground state, together with the Landau-Zener-St\"uckelberg (LZS) process which helps understand the presence of oscillatory features in our results. The LZ model is represented by a two-level system following the Hamiltonian~\cite{Landau, Zener}
\begin{equation}\label{hlandau}
\hat{H}_{LZ}=-\frac{\Delta_{0}}{2}\hat{\sigma}_{x}+\frac{\lambda_{0}-\lambda(t)}{2}\hat{\sigma}_{z}\ \ .
\end{equation}
The energy-gap between the ground state and the excited state is $\Delta(t)=\sqrt{\Delta_{0}^{2}+\left(\lambda(t)-\lambda_{0}\right)^{2}}$. At the QCP $\lambda=\lambda_{0}$, the system reaches its minimal energy-gap $\Delta=\Delta_{0}$. In this paper, we are considering the light-matter interaction $\lambda(t)=\nu t$ during the ramping up, and a similar form in the ramping down, where $\nu$ is an annealing velocity and where the system starts from its ground state, i.e. $\vert0\rangle$.  It is known from previous work on the LZ model that for a two-level system starting from its ground state at $t=-\infty$, the probability of it ending in its excited state at $t=\infty$ is given by  $ P_{LZ}=\exp\left(-\frac{\pi\Delta_{0}^{2}}{\nu}\right)$ where the ratio $\zeta=\frac{\Delta_{0}^{2}}{\nu}$ is called the adiabatic parameter. It is worth noticing that when $\nu$ is very small, we have zero probability for the system to jump to the excited state. Therefore, the limit $\nu\to 0$ corresponds to perfect adiabatic evolution. Thus the parameter $\zeta$ allows control of the probability for the system to perform either an adiabatic or diabatic transition.\\
\\
One interesting aspect of the LZ model is that much of its dynamics is determined during the short interval during which the minimum-gap is crossed. For this reason, an LZ system can be seen as analogous to a beam-splitter~\cite{HuangPRX}, since the probability $P_{LZ}$ for the system to stay in state $\vert 0 \rangle$ can be seen as equivalent to a transmission coefficient which characterizes the probability for a beam to go through a partially reflecting mirror. The analogy with the beam-splitter can be extended to one of interest to this paper, in which the evolution of $\lambda(t)$ is reversed and the critical gap is crossed again. This complete cycle is known as a Landau-Zener-St\"uckelberg (LZS) process. At the end of a LZS cycle, the probability of staying in the $\vert 0 \rangle$ state can be approximated by the following formula~\cite{HuangPRX}:
\begin{equation}\label{ProLZS}
P_{LZS}=4P_{LZ}\left(1-P_{LZ}\right)\sin\left(\theta_{12}-\Phi_{S}\right),\qquad\text{with}\quad \theta_{12}=\int_{t_{1}}^{t_{2}}\Delta(t)\text{d}t.
\end{equation}
The times $t_1$ and $t_2$ define the interval between the two crossings of the gaps. The phase $\Phi_{S}$ is called the Stokes phase and it is determined entirely by the form of the minimum gap -- hence it does not depend on the annealing  velocity $\nu$. On the other hand, the dynamical phase $\theta_{12}$ is inversely proportional to $\nu$; or equivalently, it is directly proportional to the total time of the cycle $\tau \propto 2\nu^{-1}$. As a result, Eq.~\eqref{ProLZS} implies that one should expect oscillatory behavior, with respect to $\tau$, for the probability of finishing the cycle in the same state as which the cycle was started. Such oscillatory behavior is known as {\em St\"uckelberg oscillations}~\cite{Stuckel}. Despite its simplicity, the LZ problem has found an enormous range of applications in various experimental situations. Also, some generalizations of its concepts can be performed in order  to tackle the dynamics in situations involving more than two levels~\cite{AtlandPRL, AtlandPRA, GuozhuNC}. Whether it can be extended to provide a full, formal description of the DM hysteresis results studied here, remains an open challenge.

\section{Results and discussion}\label{results}

\begin{figure}
\begin{center}
\includegraphics[scale=0.8]{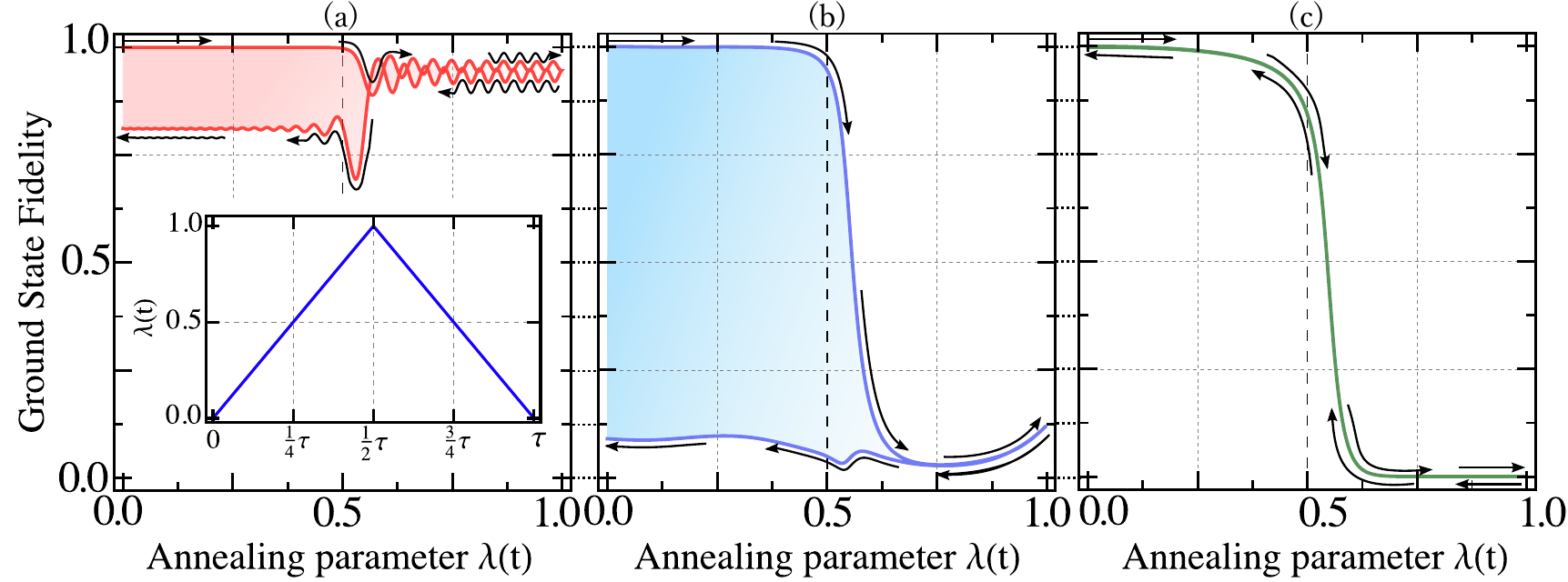}
\caption{\label{Fide}  Quantum hysteresis as measured by ground state fidelity $\vert \langle \varphi^{GS}_{Ins} \left(\lambda(t)\right)\vert \psi(t)\rangle \vert$ as a function of $\lambda(t)$ when the interaction parameter performs a cycle as specified by Eq.~\eqref{ciclo}. The DM system size is $N = 33$. The cycle time $\tau=2/\nu$ in each case is characterized by an annealing velocity $\nu$:  {\bf (a)} $Log_{2} \left(\nu\right)=-8.46$, {\bf (b)} $Log_{2} \left(\nu\right)=-4.46$, and {\bf (c)} $Log_{2}\left(\nu\right)=2.94$. Hence the annealing velocity $\nu$ increases from {\bf (a)} to {\bf (c)}. Inset in panel {\bf(a)} shows the time profile of the annealing parameter cycle $\lambda(t)$ specified by Eq.~\eqref{ciclo}.
}
\end{center}
\end{figure}

Although the LZ and LZS can serve as a guide for understanding the full numerical results of the DM, the complexity of the DM's non-equilibrium quantum dynamics involves many more than two energy levels. For any value of $\lambda(t)$, there is a set of instantaneous eigenstates $\vert\varphi_{n}\left(\lambda(t)\right)\rangle$. If $\vert\psi(t)\rangle$ represents the actual dynamical state, we can express the probability of being in the instantaneous eigenstate $n$ as follows:
\begin{equation}
P_n \left(t\right) = \left[F\left(\varphi_{n}\left(\lambda(t)\right) , \psi(t)\right)\right]^{2}
\end{equation}
where $F\left(\varphi_{n}\left(\lambda(t)\right) , \psi(t)\right)=\vert \langle \varphi_{n} \left(\lambda(t)\right)\vert \psi(t)\rangle \vert$ is the instantaneous fidelity of the $n$ state.  We consider the ground state fidelity $F(t)=\vert \langle \varphi^{GS}_{Ins} \left(\lambda(t)\right)\vert \psi(t)\rangle \vert$ as a reference quantity for characterizing the quantum hysteresis, where $\varphi^{GS}_{Ins} \left(\lambda(t)\right)$ is the ground state of the Hamiltonian that corresponds to time $t$. In Fig.~\ref{Fide} we plot the dependence of the function $F(t)$ for different values of $\nu$. Despite the fact that the finite-size DM has no true QPT, the system dynamics reveal significant differences between what would be the normal phase $\left(\lambda(t)<\lambda_c\right)$ and the superradiant one. This manifests itself in the curves since all functions $F(t)$ start to depart from their initial value of unity after a threshold near the QCP is crossed. As can be deduced from Fig.~\ref{Espectro}{\bf (b)}, the critical point of the finite-size model is slightly above the TL critical value, displaced to the right in the plot. The crossing of the phase-boundary also manifests itself in the return stage of the cycle. In general, one can see that the normal phase tends to stabilize the behavior while the superradiant phase is the one in which most changes emerge.
Very low values of $\nu$ (or equivalently, very high values of $\tau$) correspond to almost zero probability of exciting the system. Therefore the system essentially follows the instantaneous eigenstate during its entire time evolution. This regime can hence be labelled the near adiabatic limit: see Fig.~\ref{Fide}{\bf (a)}. In the other extreme of very short cycle times, the system is simply not able to respond to the change of the Hamiltonian. Hence it remains frozen in the initial state. This is the sudden quench limit: see Fig.~\ref{Fide}{\bf (c)}. In this limit, the decrease in the fidelity is due to the differences  between the instantaneous ground state and the starting one as predicted by the QPT. Hence this change is concentrated around the QCP. In between these two limits, strong memory effects occur as shown in Fig.~\ref{Fide}{\bf (b)}. The trajectories enclose an area that can be referred to as a signature of quantum hysteresis behavior. Recent experimental realizations of the DM~\cite{KlinderPNC2015} show results of such possible effects.\\
\\
\begin{figure}[h!]
\begin{center}
\includegraphics[scale=1.1]{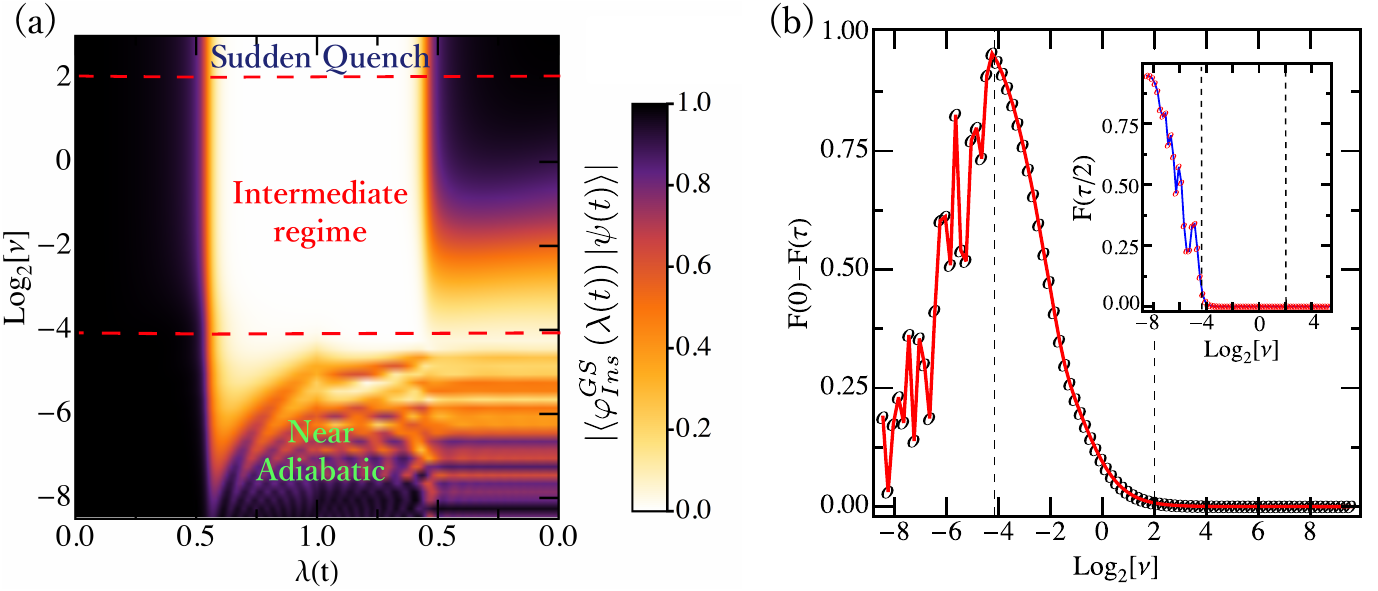}
\caption{\label{Fide2}  {\bf (a)} Quantum hysteresis profiles as measured by the ground state  fidelity $\vert \langle \varphi^{GS}_{Ins} \left(\lambda(t)\right)\vert \psi(t)\rangle \vert$ as a function of $\lambda(t)$. The dashed lines are a guide to the eye. {\bf (b)} Difference between the values of the ground state fidelity at the start time $t=0$ and end time $t=\tau$.  Inset  in Fig.~\ref{Fide2} {\bf(b)} shows the values of the ground state fidelity at the end time $t=\tau/2$. The continuous lines are a guide to the eye. In {\bf(a)} and {\bf(b)} the interaction parameter performs the cycle specified by Eq.~\eqref{ciclo}. The profile show the existence of the three regimes of annealing velocity: {\it (i)} near adiabatic, {\it (ii)} intermediate regime, and {\it (iii)} sudden quench. The DM system size is $N=33$.}
\end{center}
\end{figure}
In order to better understand the transition between the near adiabatic and sudden quench limits, we present in Fig.~\ref{Fide2}{\bf (a)} a dynamical profile of the time evolution of the ground state fidelity for a wide range of annealing velocities. It is clear that there is a non-monotonic path between the $\nu \to 0$ limit (i.e. logarithm of $\nu$ is negative) and $\nu \to \infty$ limit (i.e. logarithm of $\nu$ is positive) . In Fig.~\ref{Fide2}{\bf (b)} we are able to clearly distinguish three regimes based on the behavior of the final fidelity as a function of $\nu$. These are {\it (i)} near adiabatic, {\it(ii)} intermediate regime, and {\it (iii)} sudden quench. Complex oscillations arise as a function of $\nu$ in the near adiabatic regime. This is a many-body version of the St\"uckelberg oscillations as defined in Eq.~\eqref{ProLZS}. The amplitude of being either in the ground or the first excited state, accumulates a dynamical phase until the system returns to the QCP, at which point these two channels interfere with each other and hence form the oscillatory pattern seen in the left part of Fig.~\ref{Fide2}{\bf (b)}. As is also evidenced in Fig.~\ref{Fide}{\bf (a)}, the time interval $\tau/4 <t <3\tau/4$ (i.e. when the system is in the superradiant phase) is dominated by oscillations that tend to disappear as the annealing velocity increases. These oscillations are restricted to the superradiant phase because it is only in this interval that there is a non-negligible transition amplitude between the ground and first excited state, due to a significant change of the ground state as a function of $\lam$. The near adiabatic region is the closest to a LZS cycle in the sense that only the ground state and first excited state of the DM are significantly excited, and hence a two-level approximation is feasible. That is why both the St\"uckelberg oscillations and the superradiant phase oscillations are only relevant for slower cycles. For faster cycles, part of the evolution information leaks to higher excited states so that the simplified LZS scenario is no longer valid.\\
\\
Notwithstanding the oscillatory behavior, the near-adiabatic regime has a general tendency to show an increase in memory effects as the cycles get faster, which is evidenced by the discrepancy between the initial and final fidelities in Fig.~\ref{Fide2}{\bf (b)}. However this tendency has an upper limit, after which the intermediate regime begins. This intermediate regime is characterized by a monotonic decrease of the difference $F(0)-F(\tau)$ as the annealing velocity increases. As can be seen in the inset of Fig.~\ref{Fide2}{\bf (b)}, in this intermediate regime the system loses any ability to follow the instantaneous ground state during the superradiant phase, but somehow manages to have a finite probability of remaining in the ground state of the normal phase after the cycle is completed. This can be interpreted as the system being significantly quenched only during its passage to the superradiant phase, in a process that cannot longer be approximated as an LZ problem. In previous works, we have shown that this process is fundamentally a squeezing mechanism in both subsystems, followed by a generation of light-matter entanglement~\cite{AcevedoPRA2015,Acevedo2015NJP}. This process becomes increasingly irrelevant in terms of being able to change the initial state as $\nu$ increases in the intermediate regime, since the system has less and less time to undergo any changes. This explains the monotonic tendency toward reduced hysteresis effects as the sudden quench limit is reached.
\begin{figure}[h!]
\begin{center}
\includegraphics[scale=1.1]{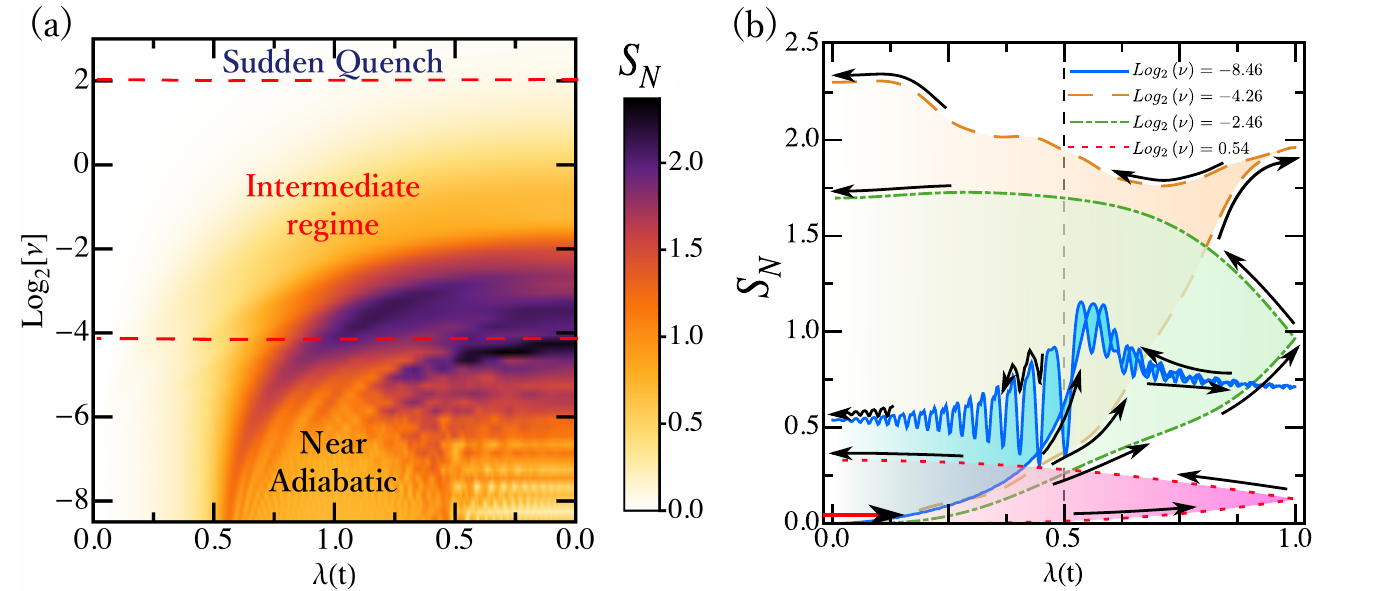}
\caption{\label{fentro}  {\bf (a)} Quantum hysteresis profile analogous to that of Fig.~\ref{Fide2}{\bf (a)} but now depicting the time evolution of the von Neumann entropy $S_N$ as defined in Eq.~\eqref{eqentro}. This quantity is an indicator of the entanglement between the light and matter subsystems. The dashed lines are a guide to the eye. {\bf (b)} Quantum hysteresis curves for $S_N$, illustrating the behavior of light-matter entanglement in the different dynamical regimes.}
\end{center}
\end{figure}

In addition to the ground state fidelity, we have analyzed the quantum hysteresis by looking at the light-matter entanglement generated by the cycle in terms of von Neumann entropy. Given a subsystem $A$, the von Neumann entropy is defined as:
\begin{equation} \label{eqentro}
S_N=-\tr{\rhoo_A\log \pap{\rhoo_A }}\:,\:\:\:\rhoo_A=\mathrm{tr}_B\pac{\ket{\psi}\bra{\psi}}
\end{equation}
where $B$ is the complementary subsystem and the total system is in a total state $\ket{\psi}$ that is pure. When the total system is in such a pure state, the entropy of subsystem $A$ is equal to the entropy of its complementary subsystem $B$, and this quantity $S_N$ is a measure of the entanglement between both subsystems. The natural choice for such a bipartition of the DM is where one subsystem is the light (i.e. the radiation mode) and the other subsystem is the matter (i.e. the set of $N$ qubits). Figures~\ref{fentro}{\bf (a-b)} show results for the von Neumann entropy for the cases discussed so far in this paper for ground state fidelity. Since the DM is a closed system (i.e. a pure quantum state with a unitary evolution), the increase of $S_N$ in each subsystem is synonymous with an irreversible interchange of information between the light and matter during the cycle, hence providing a more direct thermodynamical interpretation for the memory effects of the cycle.\\
\\
In Figs.~\ref{fentro}{\bf (a-b)}, the near adiabatic regime shows a new aspect of interest: the von Neumann entropy is not always increasing over time, which means that for slow annealing velocities, information is not always dispersing from light to matter and vice versa. Instead, there is some level of feedback for each subsystem, so that they are still able to retain some of their initial state independence. However, this feedback becomes increasingly imperfect so that at annealing velocities near the boundary with the intermediate regime, the information mixing attains maximal levels. After that, the mixing of information between light and matter is always a monotonic dispersion process, which becomes reduced as the time of interaction is reduced more and more. This establishes a striking difference between the lack of memory effects in the adiabatic and sudden quench regimes: the former's cycle comprises a large but reversible change, while the latter's cycle is akin to a very small but irreversible one. In practice, both mean relatively small changes to the initial condition -- however this is a consequence of two very different properties. This interplay between actual change and its reversibility may explain why the transition between those two regimes is more intricate that might initially have been imagined.

\section{Conclusions}\label{conclusions}

We have presented a quantum hysteresis analysis of the finite-size Dicke model (DM) in cycles that cross the quantum phase transition (QPT) from the non-interacting to the strong coupling regime and back again. In order to explore and quantify the resulting collective memory effects, we have employed the ground state fidelity and the light-matter entanglement measured through the light and matter von Neumann entropy. The former measure is more oriented to adiabatic quantum control, while the latter is more related to quantum information and quantum thermodynamics. We have identified the entire range of regimes of the cyclic dynamical process: from the adiabatic limit at small annealing velocities, to the sudden quench regime. This revealed that the transition between these two regimes is by no means a trivial one, due to an interplay between the amount of change undergone by the system as compared to the reversible character of that change. Towards the near adiabatic regime, some degree of reversibility is possible despite the system being forced to undergo large changes, which means that information can still go back and forward between the light and matter subsystems. This generates an oscillatory behavior comparable to LZS processes. By contrast towards the sudden quench regime, the information exchange between light and matter is always dispersive but gets smaller and smaller as the interaction times are reduced. These two regimes could have their own interesting applications which we leave for future exploration. In particular, the interference occurring in the near adiabatic regime could be important for spectroscopy applications, since it reveals details of the interaction during the hysteresis process. By contrast, characterization of the intermediate regime is important for quantum control, since it improves understanding of the squeezing process that precedes the dynamical generation of light-matter entanglement.

\acknowledgments{ F.J.G-R., F.J.R., and L.Q. acknowledge financial support from Facultad de Ciencias through UniAndes-2015 project \emph{Quantum control of nonequilibrium hybrid systems-Part II}. O.L.A. acknowledges support from NSF-PHY-1521080, JILA-NSF-PFC-1125844, ARO, AFOSR, and MURI-AFOSR. N.F.J. acknowledges partial support from the National Science Foundation (NSF) under grant CNS1500250 and the Air Force under AFOSR grant 16RT0367.}

\bibliography{mybib}	
\end{document}